\RequirePackage{fix-cm}
\documentclass[twocolumn]{svjour3}          % twocolumn

\smartqed  % flush right qed marks, e.g. at end of proof
\usepackage{graphicx}
\usepackage[fleqn]{amsmath}
 \newcommand{\hg}{ }
\journalname{To appear on Nonlinear Dynamics }
\begin{document}

\title{Amplitude Stochastic Response of Rayleigh Beams to Randomly Moving Loads}
%\subtitle{}

%\titlerunning{Short form of title}        % if too long for running head
\author{L. M. Anague Tabejieu \and B. R. Nana Nbendjo\and G. Filatrella \and  P. Woafo}

\authorrunning{L. M. Anague Tabejieu et al.} % if too long for running head

\institute{L. M. A. Tabejieu\and B. R. N. Nbendjo\and P. Woafo \at
              Laboratory of Modelling and Simulation in Engineering, Biomimetics and Prototypes,\\
               Faculty of Science, University of Yaounde I, P.O. Box 812, Yaounde, Cameroon. \\
              \email{nananbendjo@yahoo.com}        \\
              G. Filatrella \at
            Department of Science and Technologies, University of Sannio, Via Port' Arsa 11, I-82100 Benevento, Italy.
}

\date{March 2017}
% The correct dates will be entered by the editor

\maketitle

\begin{abstract}
We consider the problem of the nonlinear response of a Rayleigh beam to the passage of a train of forces moving with stochastic velocity.
The Fourier transform and the theory of residues is used to estimate the mean-square amplitude of the beam, while the stochastic averaging method  gives the stationary probability density function of the oscillations amplitude. The analysis shows that  the effect of the load random velocities is highly nonlinear, leading to a nonmonotonic behavior of the mean amplitude versus the intensity of the stochastic term and of the load weight. The analytic approach is also checked with numerical simulations. The effect of loads number on the
system response is numerically investigated.
% \PACS{PACS code1 \and PACS code2 \and more}
% \subclass{MSC code1 \and MSC code2 \and more}
\end{abstract}

\section{Introduction}
\label{sec:introduction}
The problem of transverse vibrations of beams subject to the passage of randomly moving loads
is of considerable practical interest in the dynamics of structures, for it has a wide range of applications in civil, mechanical and aircraft industries.
This problem has been especially studied in the context of the bridges behavior;
for example, Fryba \cite{Fryba1,Fryba2} has investigated the response of a simply supported beam using an Euler-Bernoulli model subject to a single moving force, as well as the response to continuous random forces, in particular the effects of the constant speed and damping on the beam dynamics.
Zibdeh \cite{Zibdeh} has investigated the vibrations of a simply supported elastic beam under the action of a point load moving with random non constant velocity, while the beam is also subject to axial deterministic forces.
Closed form solutions for the mean and variance of the response have also benn obtained.
To take into account the rotary and high frequency motion of beam elements, Chang \cite{Chang} proposed to treat the deterministic and random vibration analysis of a Rayleigh-Timoshenko beam on an elastic foundation with modal analysis to compute the dynamic responses of the structure (such as the displacement and bending moment) and some statistical responses (such as the mean square values of the dynamic displacement and the mean bending moment).
With the same method Argento {\it et al.} \cite{Argento} have studied the response of a rotating Rayleigh beam with different boundary conditions subject to an axially accelerating distributed surface line load.

The above mentioned works deal with bridge beams vibrations caused by a random stream of moving forces, assuming that the loads move with an exactly determined velocity.
A step further is to consider the velocities as stochastic variables \cite{Sniady1,Sniady2,Sniady3}.
If the beam is loaded by stochastically moving loads, the problem is more complicated and, generally, only numerical methods allow to retrieve the resulting vibrations \cite{Bryja}

The aim of this work is to give an analytical approach to characterize the probabilistic features of the nonlinear beam response, namely the mean-square amplitude and the probability density function due to loads moving with stochastic velocities.
The effects of the number of moving loads, the mean velocity and the intensity of the velocity disturbances on the dynamics of the beam are also analysed.
\hg {
We believe that we have made progresses in three directions: (a) Modelling the dynamic response of a Rayleigh two way beam subjected to a train of forces moving with stochastic velocity in such a manner that is, at least in the lowest order, tractable; (b) Estimating the mean square amplitude of the beams by using the Fourier transform and the theory of residues method; (c) Exploiting the stochastic averaging method to retrieve the stationary probability density function and its dependence on the intensity of the stochastic velocity. The latter result is, in our view, the most striking effect we have found.
}

The paper is organized as follows. After this Introduction, the physical model and the algorithm for numerical simulations are presented in Section {\ref{sec:model}}. In Section \ref{analysis}, the results of the Fourier analysis of the modal equation is described. Then, the results of the stochastic averaging method, the amplitude of the oscillations and the stationary probability density function of the amplitude for the noisy modal equations are retrieved in Section {\ref{sec:stationary}}.  Finally, Section {\ref{sec:conclusion}} is devoted to concluding remarks.

%%%%%%%%%%%%%%sec%%%%%%%%%%%%%%%%%%%%%%%%
\section{The Beam Model }
\label{sec:model}
This Section is devoted to the presentation of the system (Subsect. \ref{subsec:mathematical}), and to the corresponding reduced modal equations (Subsect. \ref{subsec:modal}).

\subsection{Mathematical modelling}
\label{subsec:mathematical}
 Let us consider a beam of finite length $L$, that is a nonlinear elastic structure.
In particular, our attention is paid to the geometric nonlinearities described due to the Euler-Bernoulli law, that states that the bending moment of the beam is proportional to the change in the curvature produced by the load \cite{fertis,oumbe,Duffing}.
To take into account the high frequency motion of the beam, a Rayleigh beam correction (up to the second order of the bending angle) \cite{seon} is used to refine the theory of Euler-Bernoulli beam for high frequency motion.
When the governing equation for the vertical displacement of the beam incorporates the Rayleigh term into the analysis, the correction affects the summing moments produced in the simple Euler-Bernoulli theory.

\begin{figure*}[hbtp]
    \begin{center}
    \includegraphics[width=5in,height=2.0in]{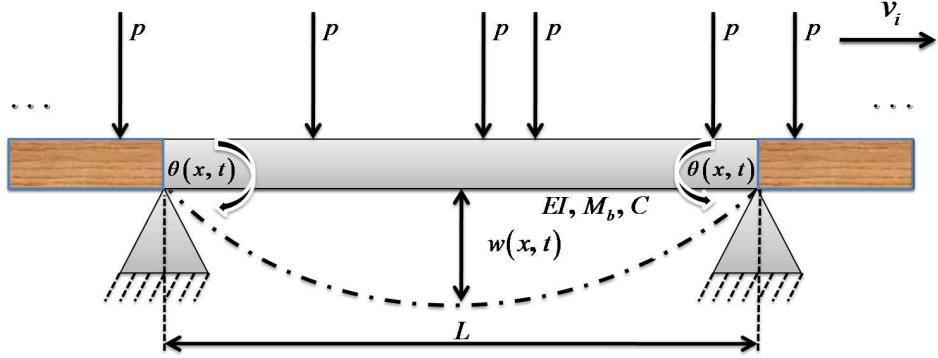}
    \caption{Sketch of a beam under stochastic moving loads. The gravitational forces are represented by arrows $P$, whose separations are not uniform, for the speeds $v_i$ are not identical.}
 \label{fig:1}
   \end{center}
   \end{figure*}

Vibrations of the beam are caused by a set of point forces of constant amplitudes {\it P}, the inter-arrival times are different, deterministic variables  $t_i$ and the forces are moving along the beam with stochastic velocities ${\it v_i}$ (see Fig.~\ref{fig:1}).

 %%%% FIGURE BEAM %%%%%%%%%%%%%%%%%%%%%%

 %%%%%%%%%%%%%%%%%%%%%%%%%%%%%%%%%%%
 Considering the classical viscous damping for the viscosity materials and Newton second law of motion, for an infinitesimal element of the beam, the equation of motion for the small deformations: $\theta\left( x, t\right)  \approx \partial w \left( x,t\right)/\partial x$ (here $w(x,t)$ is the transverse defection of the beam at point $x$ and time $t$) is obtained as

   \begin{center}
    \begin{equation}\label{eq1}
 \begin{array}{l}
 {M_b}\frac{{{\partial ^2}w\left( {x,t} \right)}}{{\partial {t^2}}} - {R_a}\frac{{{\partial ^4}w\left( {x,t} \right)}}{{\partial {t^2}\partial {x^2}}} + C\frac{{\partial w\left( {x,t} \right)}}{{\partial t}} + \\
\\    \,
 EI\frac{{{\partial ^2}}}{{\partial {x^2}}}\left[ {\frac{{{\partial ^2}w\left( {x,t} \right)}}{{\partial {x^2}}}\,\left( {1 - \frac{3}{2}\,{{\left( {\frac{{\partial w\left( {x,t} \right)}}{{\partial x}}} \right)}^2}} \right)} \right] =\\
 \\ \,
% \,\,\,\,\,\,\,\,\,\,\,\,\,\,\,\,\,\,\,\,\,\,\,\,\,\,\,\,\,\,\,\,\,\,\,\,\,\,\,\,\,\,\,\,\,\,\,\,\,\,\,\,\,\,\,\,\,\,\,\,\,\,\,\,\,\,\,\,\,\,\,\,\,\,\,\,\,\,\,\,\,\,\,\,\,\,\,\,\,\,\,\,\,\,\,\,\,\,\,\,\,\,\,\,\,\,\,\,\,\,\,\,\,\, 
= P\sum\limits_{i = 1}^{{N_v}} {{\varepsilon _i}\delta \left[ {x - {x_i}(t - {t_i})} \right]}
 \end{array}
  \end{equation}
  \end{center}

 \noindent
 with the boundary and initial conditions
 \begin{equation}\label{eq2}
 \begin{array}{l}
  \left.w\left( {x,t} \right) \right|_{x = 0, L }  = 0, \,\,\, \left. {\frac{{\partial ^2 w\left( {x,t} \right)}}{{\partial x^2 }}} \right|_{x = 0, L }  = 0.\\
 \\
 \left.w\left( {x,t} \right) \right|_{t = 0 }  = 0 \,\,,\,\left.\frac{{\partial w\left( {x,t} \right)}}{{\partial t}} \right|_{t = 0 }  = 0.
 %w\left( {0,t} \right) = w\left( {L,t} \right) = 0\,\,,\,\frac{{{\partial ^2}w\left( {0,t} \right)}}{{\partial {x^2}}} = 0\,,\,\,\frac{{{\partial ^2}w\left( {L,t} \right)}}{{\partial {x^2}}} = 0\\
 %\\
 %w\left( {x,0} \right) = 0\,\,\,,\,\,\frac{{\partial w\left( {x,0} \right)}}{{\partial t}} = 0
 \end{array}
 \end{equation}
Here $EI$ denotes the flexural rigidity of the beam, $M_b$ the beam mass per unit length, $C$ the damping coefficient, $R_a$ the transverse Rayleigh beam coefficient and $\delta[...]$ the Dirac delta function.
The position of the $i^{th}$ force at the time $t$ reads ${x_i}\left( {t - {t_i}} \right)$, where $t_i = (i-1)d/v_0$ is the deterministic arriving time of the $i^{th}$ load at the beam.
The average spacing loads is $d$ , $v_0$ the average speed of the moving loads, and $N_v$ the number of the applied loads.
Eq.~(\ref{eq1}) contains a nonlinear term due to a possible high deflection motion of the beam \cite{fertis,oumbe,Duffing}.  To facilitate a compact representation of the equations, a window function $\varepsilon_i$ is employed \cite{Nikkhoo}: ${\varepsilon _i} = 0$ when the load has left the beam and ${\varepsilon _i} = 1$ while the load is crossing the beam.

A more realistic and useful model of highway traffic loads takes into account random arrival times, in the form, for examples, of Poisson process \cite{Tung1} or renewal counting process \cite{Tung2}, to represent the vehicular traffic fluctuations.
Load fluctuations are thus assumed to be a set of point forces of constant amplitudes whose inter-arrival times are different, for the forces are moving with stochastic velocities.

Thus, let us consider that the velocities \textbf{$v_i(t- {t_i})$} are Gaussian distributed around the average speed $v_0$ \cite{Sniady1}. \hg{These velocities are a function of the difference between the time $t$ and  and the instant at which the vehicles enter the bridge, $t_i = (i-1)d/v_0$}
 \begin{equation}\label{eq3}
 \begin{array}{l}
 \frac{{d{x_i}(t- {t_i})}}{{dt}} = {v_i}(t- {t_i}) = {v_0} + {\sigma _v}{\xi _i}(t- {t_i}),\\
 \\
 0 \le {x_i}(t- {t_i}) \le L.
 \end{array}
 \end{equation}
Here $v_i(t- {t_i})$ is the stochastic velocity of the $i^{th}$ force, $\sigma_v$ its standard deviation and $\xi_i(t- {t_i})$ the velocity disturbances which it is assumed to be independent and stationary white noise random processes; i.e.
 \begin{equation}\label{eq4}
 \begin{array}{l}
  \langle {v_i}(t- {t_i}) \rangle  = {v_0}\,,\,\,\,\, \langle {\xi _i}(t- {t_i}) \rangle  = 0,\\
 \\
  \langle {\xi _i}(t- {t_i}){\xi _j}(t- {t_i}) \rangle  = 0\,\,\,\,\,\,{\rm{for }}\quad i \ne j,\\
 \\
  \langle {\xi _i}(t- {t_i}){\xi _i}(t - {t_i} + \zeta) \rangle  = \,\,\gamma _v^2\delta (\zeta ).
 \end{array}
 \end{equation}
 The brackets $ \langle ... \rangle $ denote the time average and  $\gamma_v = v_0 \sigma_v$.

In the presence of the stochastic term there is a finite probability that the fluctuations produce a negative velocity, that become sizeable when fluctuations $\sigma_v$ are comparable with the average speed $v_0$.
We do not exclude such negative velocities, that we interpret as a vehicle moving in the opposite direction, thus assuming that the beam represents in fact a two-way bridge.

 \subsection{Modal equations}
 %%%%%%%%%%%%%% subsec:moda l%%%%%%%%%%%%%%%%%%%%%%
 \label{subsec:modal}
 If one takes into account the boundary conditions, the transversal deflection $w(x, t)$ for the simply-supported beam can be represented in a series form as
  \begin{equation}\label{eq5}
 w\left( {x,t } \right) = \sum\limits_{n = 1}^\infty  {{q_n}\left( {t } \right)} \sin \left( {\frac{{n\pi x}}{L}} \right)
  \end{equation}

Here, $q_n\left( t\right)$ is the amplitude of the $n^{th}$ mode, and $\sin\left( n \pi x/L\right)$ is the solution of the eigen value problem, that depends on the boundary conditions of the free oscillations of the beam.
It is convenient to adopt the following dimensionless variables:
  \begin{equation}\label{eq6}
  {\chi _n} = \frac{{{q_n}}}{r}\,\,\,,\,\,\tau  = \frac{{\tilde vt}}{L}\,\,\,,\,\,v = \frac{{{v_0}}}{{\tilde v}}\,\,
  \end{equation}
 The equivalent stochastic dimensionless modal equation is obtained substituting Eq.~(\ref{eq3}) and  Eq.~(\ref{eq5}) into Eq.~(\ref{eq1}) and \hg{considering the first mode only. We have limited the analysis to the first mode, inasmuch the first mode of vibrations is expected to carry most of the energy, and therefore one hopes that it could suffice to obtain a first estimate of the system behavior. If we indicate with $W_{i}(\tau - \tau_i)$ a unit Wiener stochastic process:}

   \begin{center}
    \begin{equation}\label{eq7}
 \begin{array}{l}
 \ddot \chi (\tau ) +\lambda \dot \chi (\tau ) + \chi (\tau ) + \beta {\chi ^3}(\tau ) = \\
=  \Gamma \sum\limits_{i = 1}^{{N_v}} {{\varepsilon _i}\sin \left[ {\Omega \tau  + \gamma W_{i}\left( \tau - \tau_i \right)} \right]}
 \end{array}
  \end{equation}
  \end{center}

with
 \begin{equation}\label{eq8}
 \begin{array}{l}
 \Omega  = \frac{{\pi {v_0}}}{{\tilde v}}\,,\,\,\,\Gamma  = \frac{{2P{L^3}}}{{rEI{\pi ^4}}}\,\,\,\,,\,\,\,\beta  =  - \frac{3}{8}{\left( {\frac{{\pi r}}{L}\,} \right)^2}\,\,,\,\,\,
\\
\,
\\
\lambda  = \frac{{C{L^3}}}{{{\pi ^2}\sqrt {EI\left( {{L^2}{M_b} + {R_a}{\pi ^2}} \right)} }}
\,\,\,\,
  {R_a} = {M_b}{r^2}\,\,,\,\,\,\,
\\
\, \\
\tilde v = {\pi ^2}\sqrt {\frac{{EI}}{{{L^2}{M_b} + {R_a}{\pi ^2}}}} \,\,,\,\,\,r = \sqrt {\frac{I}{S}} \,\,,\,\,\,\,\,\,\gamma  = \pi\sigma_v \,\,,\,\,\,\,
\\
\, \\
 \tau_i = \frac{{\tilde v  t_i}}{L}.
 \end{array}
 \end{equation}
\hg{The ansatz (\ref{eq5}) of the modal analysis is but the simplest method to deal with this intrinsically nonlinear problem, for the nonlinear terms of Eq.~(\ref{eq1}) only results in the cubic nonlinearity of Eq.~(\ref{eq7}). However, the method is capable both to retain (to a limited and approximated extent) the essential nonlinearity of the full partial differential equation, and at the same time allows for an analytical treatment, as we shall see in the following. Also, the roughness of the method calls for detailed numerical simulations to validate the analysis. As we shall see in the following, numerical simulations support the usefulness of the method.}

Eq.~(\ref{eq7}) amounts to a stochastic Duffing oscillator which describes the unbounded or catastrophic motion of the beam.  The catastrophic behavior of the beam is related to the configuration of the potential of the system, as described in details in Ref. \cite{nana}

%%%%%%%%%%%%%%%sec%%%%%%%%%%%%%%%%%%%%%%%%
\section{Modal Analysis}
\label{analysis}
It has been shown by Iwankiewicz and \'{S}niady \cite{Iwankiewicz} that the influence of the free vibration (${\varepsilon _i} = 0$) on the probabilistic characteristics of the structure response is negligibly small when the speed of the moving forces is below $130$ km/h (about $36$ m/s).
Therefore, for simplicity, it is assumed in the first case (${\varepsilon _i} = 0$) that the dynamic response function is equal to zero ($\chi (\tau ) = 0$) and for the second case (${\varepsilon _i} = 1$) the response function is calculated from the equation (considering the case of a single moving load):

  \begin{equation}\label{eq9}
  \ddot \chi (\tau ) + \lambda \dot \chi (\tau ) + \chi (\tau ) + \beta {\chi ^3}(\tau ) = \Gamma \sin \left[ {\Omega \tau  + \gamma W\left( \tau  \right)} \right]
  \end{equation}

  \noindent It is observed that the right hand side of Eq.(\ref{eq9}) is a harmonic function with constant amplitude and random phases (mathematically equivalent to frequency fluctuations of a nonmonochromatic drive \cite{Filatrella02}), and therefore it amounts to a bounded or sine-Wiener noise $\eta (\tau )$ \cite{Bobryk05} , whose covariance is given by

 \begin{equation}\label{eq10}
\begin{array}{l}
 {C_\eta }(\tau ,\tau') =  \langle \eta (\tau )\eta (\tau') \rangle  = \\
\, \\
= \frac{{{\Gamma^2}}}{2}\exp \left( { - \frac{{{\gamma ^2} {|\tau  - \tau'|}}}{2}} \right)\cos \Omega (\tau - \tau').
\end{array}
 \end{equation}
\noindent The response of the stochastic Eq.~(\ref{eq9}) is obtained using the Fourier transform associated with the residues theorem \cite{Datta}, as will be shown in the following.
\subsection{Fourier Analysis}
\label{Fourier}
The equivalent linearization of  Eq.(\ref{eq9}) is
\begin{equation}\label{eq11}
\ddot \chi (\tau ) + \lambda \dot \chi (\tau ) + \omega^2\chi (\tau ) = \eta (\tau )
\end{equation}
where
\begin{subequations}\label{eq12}
\begin{align}
\label{eq12a}
\omega^2 & = 1 + \frac{3}{4}\beta {A^2}, \\
\label{eq12b}
\eta (\tau ) &= \Gamma \sin \left[ {\Omega \tau  + \gamma W(\tau )} \right]
\end{align}
\end{subequations}

\noindent and $A$ is the root mean-square displacement.
Using the Fourier Transformations one obtains

\begin{equation}\label{eq13}
\begin{array}{l}
TF\left( {\chi (\tau )} \right)\, = \chi (\omega ')\,,\,\,\,\,\,\,\,\,\\
\\
TF\left( {\eta (\tau )} \right)\, = \eta (\omega ')\,,\,\,\\
\\
TF\left( {{\chi ^{(n)}}(\tau )} \right)\, = {(i\omega ')^n}TF\left( {\chi (\tau )} \right)
\end{array}
\end{equation}

\noindent (here  $i^2 =  - 1\,\,\,$ and $TF(.)$ is the Fourier transform operator).
The solution of Eq.~(\ref{eq11}) in Fourier space reads

\begin{equation}\label{eq14}
\chi (\omega ') = \frac{{\eta (\omega ')}}{{{\omega^2} - {{\omega '}^2} + i\lambda \omega '}}
\end{equation}

\noindent From Eq.~(\ref{eq10}), one have

\begin{equation}\label{eq15}
 \langle \eta (\omega ')\,\eta (\omega '') \rangle  = \delta \left( {\omega ' + \omega ''} \right){S_\eta }(\omega ')
\end{equation}

\noindent where ${S_\eta }(\omega ')$ is the spectral density of the noise $\eta(\tau)$ defined by

\begin{equation}\label{eq16}
{S_\eta }(\omega ') = \frac{{{{\left( {\Gamma \gamma} \right)}^2}}}{{2\pi }}\left[ {\frac{1}{{4{{\left( {\omega ' - \Omega } \right)}^2} + {\gamma ^4}}} + \frac{1}{{4{{\left( {\omega ' + \Omega } \right)}^2} + {\gamma ^4}}}} \right]
\end{equation}
Hence,
\begin{equation}\label{eq17}
\begin{array}{l}
 \langle \chi (\omega ')\chi (\omega '') \rangle \,\,\,\, = \,\,\,\frac{\begin{array}{l}
 \langle \eta (\omega ')\eta (\omega '') \rangle \\

\end{array}}{{\left( {{\omega ^2} - {{\omega '}^2} + i\lambda \omega '} \right)\left( {{\omega ^2} - {{\omega ''}^2} + i\lambda \omega ''} \right)}}\\
\\
\,\,\,\,\,\,\,\,\,\,\,\,\,\,\,\,\,\,\,\,\,\,\,\,\,\,\,\,\,\,\,\,\,\,\,\,\, = \frac{\begin{array}{l}
\delta \left( {\omega ' + \omega ''} \right)\,\,{S_\eta }(\omega ')\\

\end{array}}{{\left( {{\omega ^2} - {{\omega '}^2} + i\lambda \omega '} \right)\left( {{\omega ^2} - {{\omega ''}^2} + i\lambda \omega ''} \right)}}
\end{array}
\end{equation}
The mean-square amplitude can be calculated as
\begin{equation}\label{eq18}
{A^2} =  \langle {\chi ^2}(\tau ) \rangle \,
\end{equation}
It follows from the above definitions that:
\begin{equation}\label{eq19}
\begin{array}{l}
{A^2}\,\,\, = \,\int {\frac{{d{\omega _1}}}{{2\pi }}} \frac{{d{\omega _2}}}{{2\pi }} \prec \chi ({\omega _1})\chi ({\omega _2}) \succ {e^{i\left( {{\omega _1} + {\omega _2}} \right)\tau }}\\
\\
 = \int {\frac{{d{\omega _1}}}{{2\pi }}} \frac{{d{\omega _2}}}{{2\pi }}\frac{{\delta \left( {{\omega _1} + {\omega _2}} \right){S_\eta }({\omega _1}){e^{i\left( {{\omega _1} + {\omega _2}} \right)\tau }}}}{{\left( {{\omega ^2} - \omega _1^2 + i\lambda {\omega _1}} \right)\left( {{\omega ^2} - \omega _2^2 + i\lambda {\omega _2}} \right)}}\\
\\
= \,\frac{{{{\left( {\Gamma \gamma} \right)}^2}}}{{2\pi }}\int {\frac{{d{\omega _1}}}{{2\pi }}\frac{{2{\gamma ^4} + 8\left( {\omega _1^2 + {\Omega ^2}} \right)}}{{\left[ {4{{\left( {{\omega _1} - \Omega } \right)}^2} + {\gamma ^4}} \right]\left[ {4{{\left( {{\omega _1} + \Omega } \right)}^2} + {\gamma ^4}} \right]\left[ {{{\left( {{\omega ^2} - \omega _1^2} \right)}^2} + {\lambda ^2}\omega _1^2} \right]}}}  \\
\\
\,\,\,\,\, = \,\frac{{{{\left( {\Gamma \gamma} \right)}^2}}}{{4\pi \lambda }}.\frac{{\left[ {{\gamma ^4} + 4\left( {{\omega ^2} + {{\left( {\lambda  - \Omega } \right)}^2}} \right)} \right]}}{{{\omega ^2}\left[ {{\gamma ^8} + 16{{\left( {{\omega ^2} + \left( {\lambda  - \Omega } \right)\Omega } \right)}^2} + 4{\gamma ^4}\left( {{\lambda ^2} + 2\left( {{\omega ^2} + {\Omega ^2} - \lambda \Omega } \right)} \right)} \right]}}
\end{array}
\end{equation}
This is the main result of this part: the dependence of the oscillations amplitude upon the beam parameters (damping $\lambda$, natural frequency $\omega_0 = 1$ and nonlinear component $\beta$) and of the loads traffic (loads weights intensity $\Gamma$, average velocity $\Omega$, and velocity fluctuation intensity  $\gamma$).
In fact substituting Eq.~(\ref{eq12a}) into Eq.~(\ref{eq19}), the amplitude can be found from the roots of
\begin{equation}\label{eq20}
\Theta _4  A^8 + \Theta _3A^6 + {\Theta _2}{A^4} + \Theta _1  A^2 = \Theta _0
\end{equation}
with

\begin{subequations}
\label{eq21}
%\begin{array}{l}
\begin{align}
%\left \{ \begin{array}{ll}
 %0
\Theta _0  = {\Gamma ^2}\left[ {{\gamma ^6} + 4{\gamma ^2}\left( {1 + {{\left( {\lambda  - \Omega } \right)}^2}} \right)} \right]  \hspace{0cm} \\
%1
%\begin{split}
\Theta _1  = 4\pi \lambda  \times \hspace{0cm}   \nonumber \\
\times  \left[ {{\gamma ^8} + 4\left( {{\lambda ^2} + 2\left( {1 + {\Omega ^2} - \lambda \Omega } \right)} \right){\gamma ^4} - \frac{{3\beta {\Gamma ^2}}}{{4\pi \lambda }}{\gamma ^2}} \right]+  \nonumber \\
 64\pi \lambda \Omega    \left[ {2\lambda  + \Omega \left( {{\Omega ^2} + {\lambda ^2} - 2 - 2\lambda \Omega } \right)} \right]  \\
%\end{split}  \\
%2
\Theta _2 = 3\lambda \pi \beta \left[ {{\gamma ^8} + 4\left( {4 + {\lambda ^2} + 2\Omega \left( {\Omega  - \lambda } \right)} \right){\gamma ^4}} \right]  +\nonumber \\
  48\lambda \pi \beta  \left[ {3 + {\Omega ^4} + \Omega \left( {\Omega {\lambda ^2} - 2\lambda {\Omega ^2} + 4\left( {\lambda  - \Omega } \right)} \right)} \right]\\
%3
\Theta _3 = 18\lambda \pi \beta ^2\left[ \gamma ^4 + 6 + 4\Omega \left( \lambda  - \Omega  \right) \right]   \\
%4
\Theta _4 =  27\lambda \pi \beta ^3  \hspace{0cm}
\end{align}
\end{subequations}

 \subsection{ Algorithm of stochastic numerical simulations}
 \label{subsec:algorithm}
 To numerically treat the random process of the nonlinear  Eq.~(\ref{eq9}), the stochastic Fourth-order Runge Kutta (RK) algorithm developed by Kasdin \cite{Kasdin} is used.
 Introducing the new variables {\it y($\tau$)} and {\it z($\tau$)}, the system (\ref{eq9}) can be rewritten as
 \begin{equation}\label{eq22}
\left\{ \begin{array}{l}
\dot \chi(\tau ) = y(\tau )\\
\\
\dot y(\tau ) =  - \lambda y(\tau ) - \chi (\tau ) - \beta {\chi ^3}(\tau ) + \Gamma \sin \left( {z(\tau )} \right)\\
\\
\dot z(\tau ) = \Omega  + \gamma \xi (\tau )
\end{array} \right.
 \end{equation}
 The RK solution for $\chi_k$, $y_k$ and $z_k$ are given by the following set of equations:
 \begin{equation}\label{eq23}
 \begin{array}{l}
 {\chi _{k + 1}} = {\chi _k} + {\alpha _1}{K_1} + {\alpha _2}{K_2} + {\alpha _3}{K_3} + {\alpha _4}{K_4}\\
 {y_{k + 1}} = {y_k} + {\alpha _1}{L_1} + {\alpha _2}{L_2} + {\alpha _3}{L_3} + {\alpha _4}{L_4}\\
 {z_{k + 1}} = {z_k} + {\alpha _1}{M_1} + {\alpha _2}{M_2} + {\alpha _3}{M_3} + {\alpha _4}{M_4}\\
 \\
 {k_1} = {y_k}\,\Delta \tau \\
 {L_1} = \left[ { - \lambda {y_k} - {\chi _k} - \beta \chi _k^3 + \Gamma \sin \left( {{z_k}} \right)} \right]\Delta \tau \\
 {M_1} = \Omega \Delta \tau  + {\zeta _1}\\
 \\
 {K_j} = \left[ {{y_k} + {Y_j}} \right]\Delta \tau \\
 {L_j} = [  - \lambda \left( {{y_k} + {Y_j}} \right) - \left( {{\chi _k} + {X_j}} \right) - \beta {{\left( {{\chi _k} + {X_j}} \right)}^3} + \\
+ \Gamma \sin \left( z_k + Z_j \right) ]\Delta \tau \\
 {M_j} = \Omega \,\Delta \tau  + {\zeta _j}
 \end{array}
 \end{equation}
 where
 \begin{equation}\label{eq24}
\begin{array}{l}
{X_j} = \sum\limits_{i = 1}^{j - 1} {{a_{j - 1}}{K_i}\,\,,\,\,\,} {Y_j} = \sum\limits_{i = 1}^{j - 1} {{a_{j - 1}}{L_i}\,\,,\,\,{Z_j} = \sum\limits_{i = 1}^{j - 1} {{a_{j - 1}}{M_i}\,,\,\,\,\,} \,} \\
\,\,\\
{\zeta _1} = {r_1}\sqrt {2{q_1}\gamma \Delta \tau } \,\,,\,\,\,{\zeta _j} = {r_j}\sqrt {2{q_j}\gamma \Delta \tau } \,\,;\,\,j = 2,\,3,\,4
\end{array}
 \end{equation}
 This numerical method can be extended to simulate the case of several moving loads. Here, $r_n$  ({\it n=1,...,4}) are the Gaussian white noise from random numbers $x_m $ ({\it m=1,...,8}) by using the Box -Mueller algorithm \cite{Knuth}.
 The step $\Delta \tau$ here used is $\Delta \tau = 0.001$ and the results are averaged over $100$ realizations. The coefficients (see Table \ref{Table1}) $\alpha_i$, $q_i$ and $a_{ji}$ are chosen in the deterministic case to ensure that $\chi_k$, $y_k$ and $z_k$ simulate the solution $\chi(\tau)$, $y(\tau)$ and $z(\tau)$ with error of order $\Delta {\tau ^5}$. That is,
 \begin{equation}\label{eq25}
\begin{array}{l}
 \chi(\tau_k)= \chi_k + O\left( \Delta {\tau ^5}\right) ,\,\,y(\tau_k)= y_k + O\left( \Delta {\tau ^5}\right),\\
  z(\tau_k)= z_k + O\left( \Delta {\tau ^5}\right)
\end{array}
 \end{equation}
 These coefficients are given in Table \ref{Table1}.\par
\begin{table*}
 \begin{tabular}{llll}
  \hline \rule[-2ex]{0pt}{5.5ex}  Coefficient&  \qquad Value&  Coefficient& \qquad\qquad Value\\
  \hline \rule[-2ex]{0pt}{5.5ex}  \begin{tabular}{l}
  $\alpha_1$ \\
   $\alpha_2$ \\
  $\alpha_3$ \\
  $\alpha_4$ \\
  $a_{21}$ \\
  $a_{31}$ \\
  $a_{32}$\\
  \end{tabular} &
  \begin{tabular}{l}
   $0.25001352164789$ \\
  $0.67428574806272$ \\
  $-0.00831795169360$ \\
  $0.08401868181222$ \\
  $0.66667754298442$ \\
  $0.63493935027993$ \\
  $0.00342761715422$ \\
   \end{tabular} &
   \begin{tabular}{l}
    $a_{41}$ \\
    $a_{42}$\\
    $a_{43}$ \\
    $q_{1}$ \\
    $q_{2}$ \\
    $q_{3}$ \\
    $q_{4}$ \\
  \end{tabular} &
  \begin{tabular}{l}
   $-2.32428921184321$ \\
   $2.69723745129487$  \\
   $0.29093673271592$  \\
   $3.99956364361748$  \\
   $1.64524970733585$  \\
   $1.59330355118722$  \\
   $0.26330006501868$  \\
  \end{tabular} \\
  \hline
  \end{tabular}\par
  \qquad \par
\caption{ { Coefficients of stochastic Fourth- order Runge Kutta.} }
 % \qquad \par
\label{Table1}
\end{table*}

 \subsection{Numerical Analysis}
 \label{subsec:analysis}
To check the validity of the analytical estimates, we have compared some analytical results with numerical simulations.
 The physical parameters of the beam used throughout the paper are: $L = 34.0 { m}$, $E = 3.0 \times {10^{10}}N/{m^2}$, $I = 3.07{m^4}$, ${M_b} = 11400.0 {K_g}/m$, $C = 350.5 N.S/m$, $S=0.02 {m^2}$, and the dimensionless parameters are: $\beta  = {\rm{ - 0}}{\rm{.52}}$, $\lambda  = {\rm{0}}{\rm{.00078}}$, $\Gamma  = {\rm{0}}{\rm{.0022}}$.

\begin{figure}[hbtp]
           \begin{center}
            \includegraphics[width=2.38in,height=2.3in]{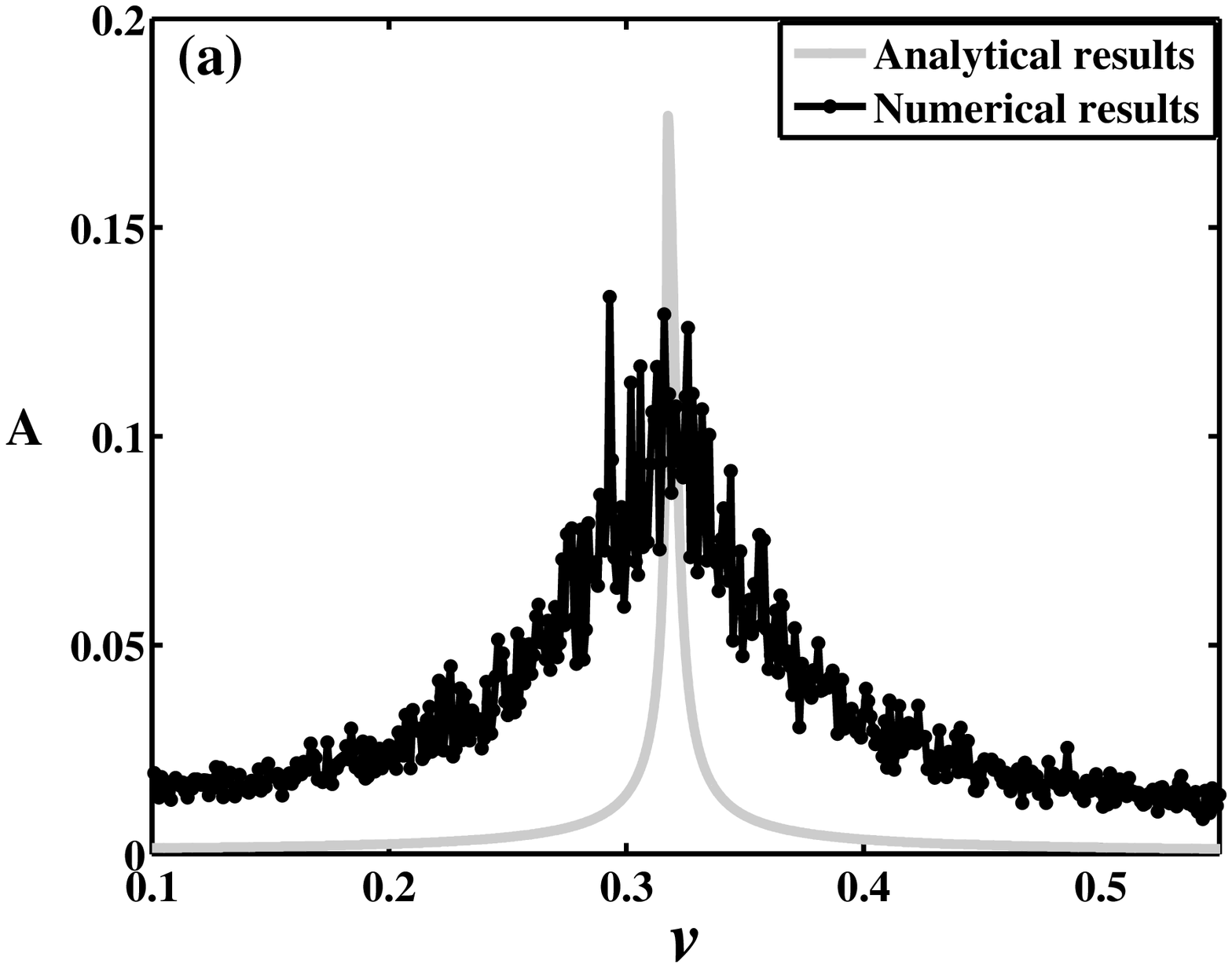}
             \includegraphics[width=2.38in,height=2.3in]{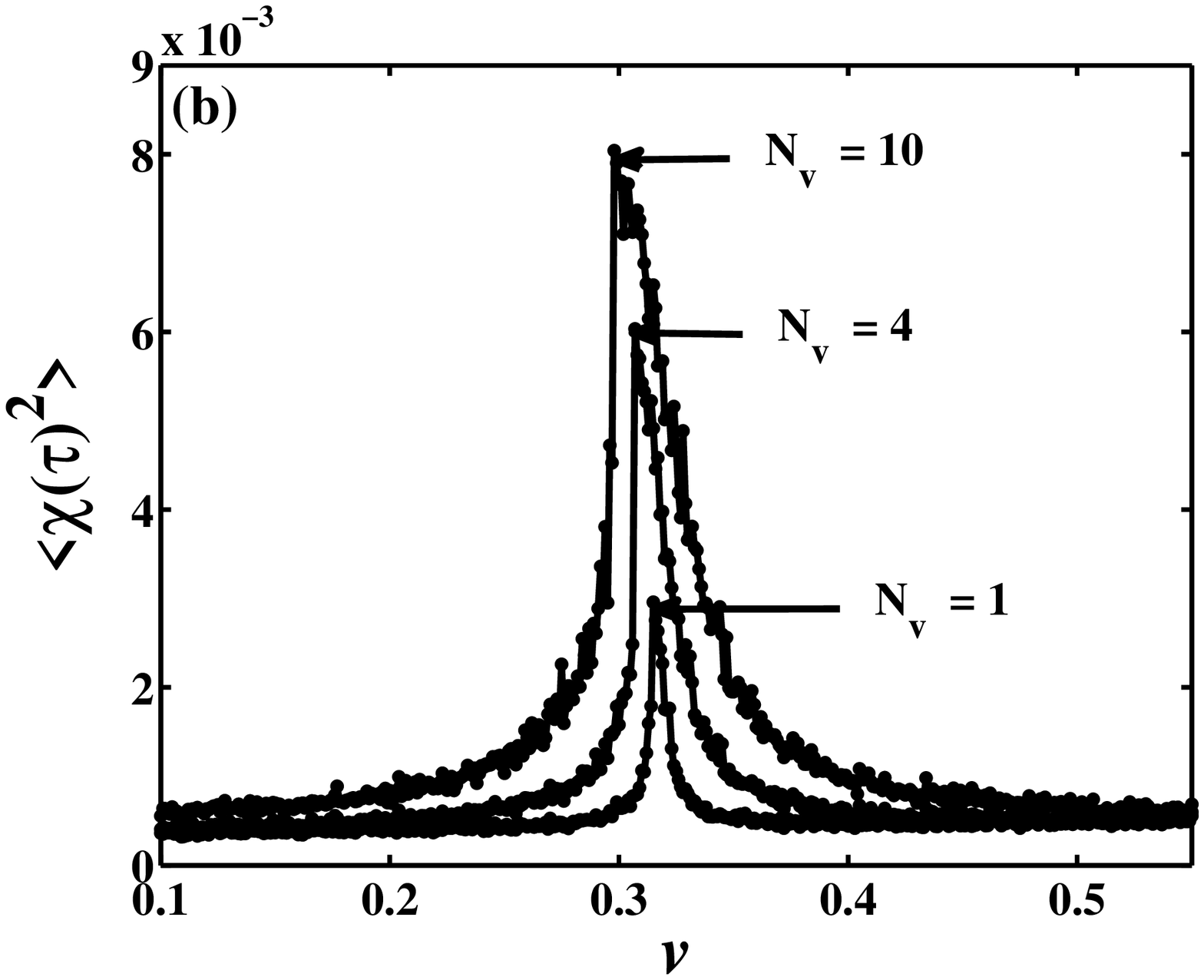}
       \caption{
(a) Comparative analysis of mean-square response through the stochastic average analysis (Eq.~ (\ref{eq20}), light gray line) and numerical simulations of the full model Eq.~ (\ref{eq9}) (black line with circle) for a single load ($N_v=1$) and for noise intensity $\gamma = 0.1$.
(b) Numerical simulations of the mean square response of a different number of loads  $N_v$ for $\gamma = 0.001$.
The other dimensionless parameters read: $\beta  = \rm{ - 0.52}$, $\lambda  = \rm{0.00078}$, $\Gamma  = \rm{0.0022}$.
}
       \label{fig:2}
           \end{center}
        \end{figure}

The mean response of the amplitude of the oscillators, as estimated with the modal equations method, Eq.~(\ref{eq20}), compared with numerical simulations of Eq.~(\ref{eq9}) obtained through the stochastic RK algorithm is displayed in Fig.~\ref{fig:2}(a). The Figure demonstrates that there is a resonant velocity $v \simeq 0.32$ where the amplitude of the oscillations increases, in both the theoretical prediction and the numerical simulations. 

\hg{  The analysis of a single load cannot be extended to the multiple vehicles case described by Eq.(\ref{eq7}),  inasmuch the superposition principle of a linear system is not established for the nonlinear dynamics system (\ref{eq7}). 
As the present analysis cannot be rigorously extended to multiple load cases, the limits of validity of the extension can only be numerically verified, as shown in Fig. \ref{fig:2}(b). 
In the Figure it is observed that the }
resonance is made more pronounced by the increase of the number of loads $N_v$. 
 Another interesting feature of Fig. \ref{fig:2}(b) is the reentrant behavior. For some values below the resonant velocity $v=0.32$ there are as many as three different amplitudes that correspond to the same speed.
  It is interesting to notice that the peculiar behavior is observed in both the analytical treatment for the noise value $\gamma = 0.1$, see Fig.~ \ref{fig:2}(a).
At this level fluctuations are comparable with the resonant speed $v_0=0.32$ in Fig.~\ref{fig:2}(a).
Physically, it corresponds to rare but sizeable negative velocities.
Instead, in Fig.~\ref{fig:2}(b) fluctuations are so small that it practically represents the case of a one way bridge.
We conclude that the analytical treatment captures the main effect, as the agreement is fairly good and the resonance is correctly captured by the analysis in a wide range of parameters.

  %%%%%%%%% FIGURE Amplitude vs velocity %%%%%%%%%%%%%%%%%%%
 %%%%%%%%%%%%%%%%%%%%%%%%%%%%%%%%%%%%%%%%%%%%%

\begin{figure*}[hbtp]
    \begin{center}
    \includegraphics[width=4.8in,height=2.7in]{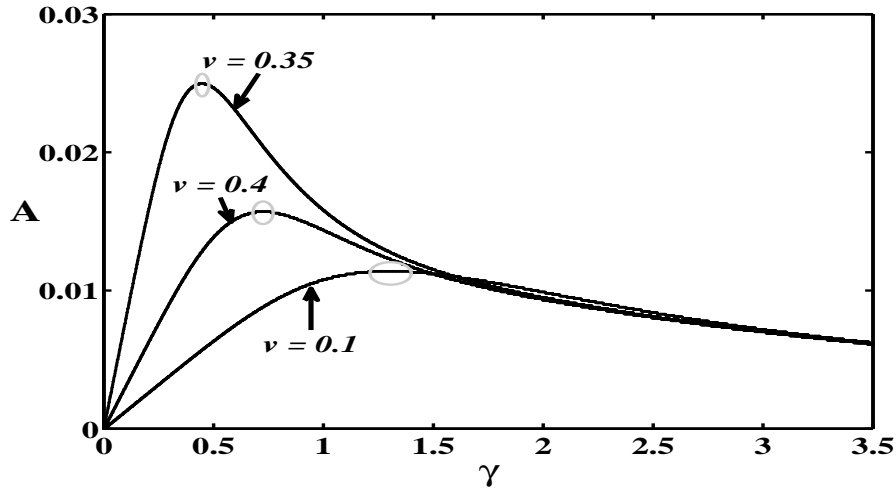}
    \caption{Influence of the intensity of the stochastic velocity on the mean amplitude of vibration for different values of the mean velocity $v$ in the averaged system Eq.(~\ref{eq9}).
The other parameters are the same as in Fig.\ref{fig:2}.}
    \label{fig:3}
   \end{center}
   \end{figure*}

The reentrant behavior is further analyzed in Fig. \ref{fig:3}, where the mean-square amplitude as a function of the noise intensity for different values of the loads speed $\it v$ is displayed.
 In Fig. \ref{fig:3} the peak of the oscillations, marked by an oval, occurs at relatively high noise values ($\gamma \simeq 1.4$)  for low speed ($v=0.1$) speeds, decreases at lower noise levels ($\gamma \simeq 0.4$) for higher speed ($v=0.35$), and moves again to higher fluctuations ($\gamma \simeq 0.6$) with a further increase of the speed ($v=0.4$).  Thus, not only the amplitude has a maximum response for a finite value of the noise intensity, but also the peak position exhibits a nontrivial behavior. In general these features can be ascribed to the peculiar character of bounded noise in model Eq.~(\ref{eq9}) \cite{Li12}.
We note that the amplitude decrease occurs for fluctuation levels that are comparable or higher than the average speed, and therefore they basically describe a two way bridge.

 There are the values of $\gamma$ for which the amplitude is almost constant at the maximum; a further increase of the noise amplitude $\gamma$ leads to a decrease of the amplitude (see Fig.~\ref{fig:3}).

  %%%%%%%%% FIGURE Amplitude vs NOISE intensity%%%%%%%%%%%%%%%%
   %%%%%%%%%%%%%%%%%%%%%%%%%%%%%%%%%%%%%%%%%%%
    %%%%%%%%% FIGURE Amplitude vs LOAD %%%%%%%%%%%%%%%%%%%
            \begin{figure}[hbtp]
             \begin{center}
             \includegraphics[width=2.38in,height=2.3in]{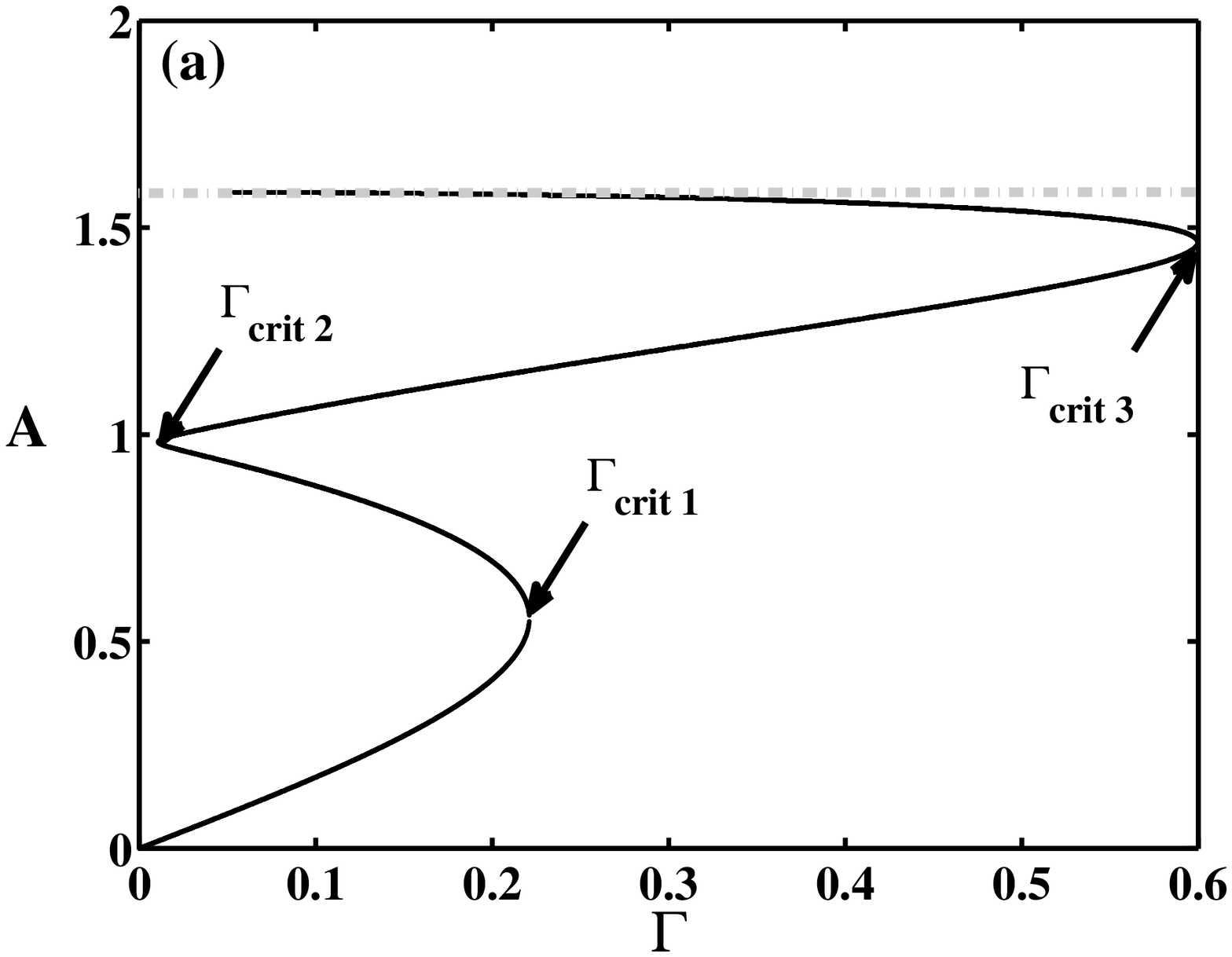}
             \includegraphics[width=2.38in,height=2.3in]{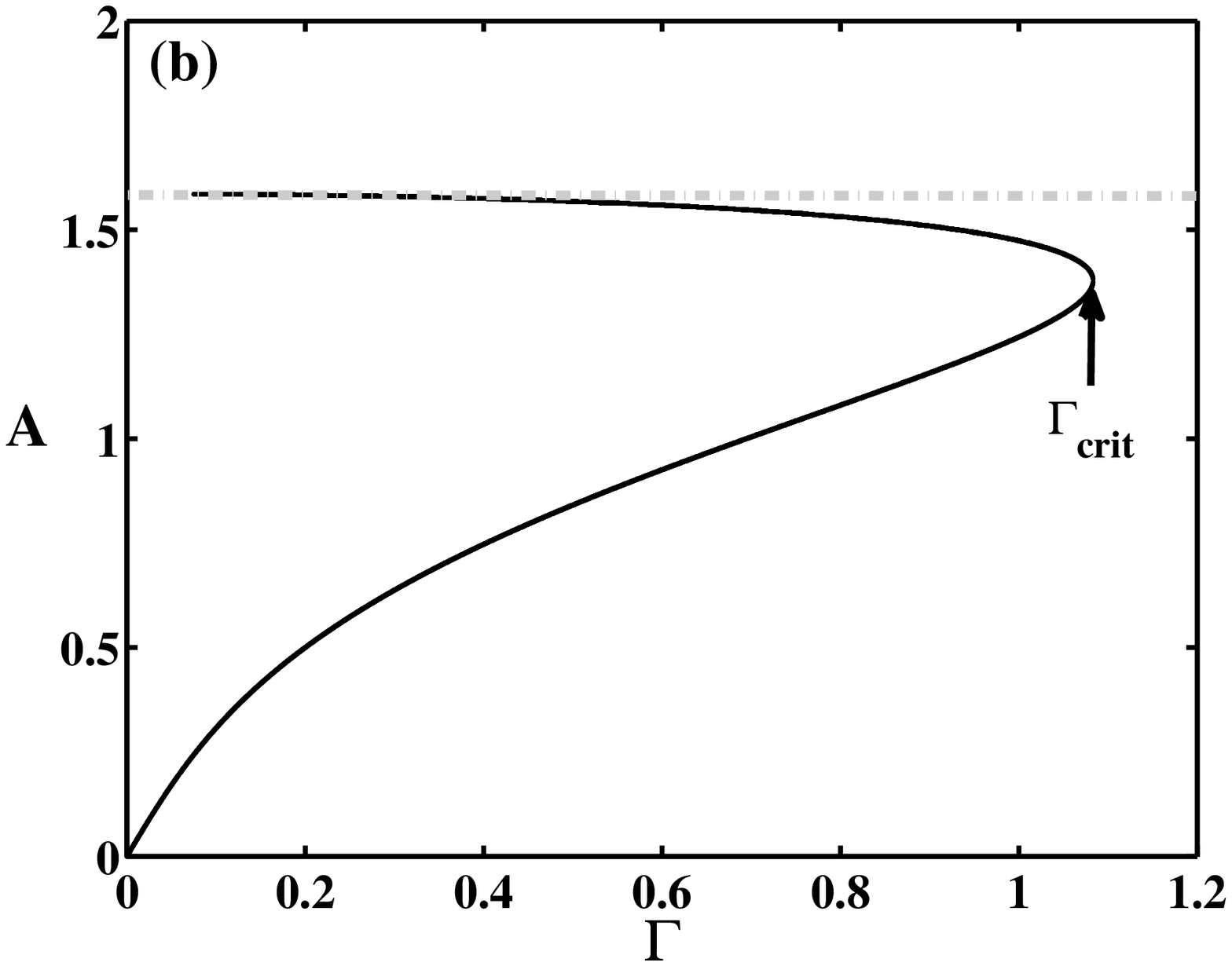}
             \caption{The mean square amplitude versus the weight load for (a) ${\it v} = 0.25$ and (b) ${\it v} = 0.35$.  It is assumed here that,  $\gamma = 0.1$.
The other parameters are the same as in Fig.\ref{fig:2}.}
             \label{fig:4}
            \end{center}
            \end{figure}
   %%%%%%%%%%%%%%%%%%%%%%%%%%%%%%%%%%%%%%%%%%

Beside the reentrant effect, it is noticed that the noise intensity tends to broad the resonance, and the number of loads increases both the mean square amplitude and the width of the peak.
   These features are naturally expected on physical grounds; more surprising is the effect of the single load weight on the mean square amplitude, as shown in Fig.~\ref{fig:4}.
    Increasing the load of each moving weight, one observes that the mean amplitude also shows a reentrant behavior with multiple solutions below the resonance speed  (Fig.~\ref{fig:4}).
   It is significant to note that when the speed is below the resonance, one observes critical values of the weight load where we have a bifurcation of the system.
   On a general ground this could be again attributed to the presence of bounded noise, that can produce bimodal distributions \cite{Bobryk05}.
   More specifically, the effect can be related to the special features of the bounded noise spectral density (\ref{eq16}), combined with the nonlinear amplitude-frequency relation of Eq.(\ref{eq20}).
   These features give rise to the intricate effects above described, inasmuch the spectral content determined by Eq.~(\ref{eq16}) is, at variance with the  uniform case of white noise, peaked around some frequencies.
   Thus, the frequency of the peaks depends upon the noise intensity: increasing  the noise intensity the harmonic content is altered.
   This is the physical origin of the reentrant behavior observed in Fig.~\ref{fig:3}: when the noise intensity is increased it first moves the harmonic content towards the resonance of the beam, increasing the amplitude of the oscillations. The effect, however, saturates and eventually the term at the denominator $\gamma$ dominates.

In Figs. \ref{fig:2},\ref{fig:3},\ref{fig:4} the mean amplitude of the oscillations of the beam has been displayed. A more complete description in terms of the full distribution is presented in the following Section, where the probability distribution of the representative coordinate $\chi$ is analysed.

\section{Stationary probability distribution of a catastrophic monostable system}
\label{sec:stationary}
\begin{figure*}[hbtp]
     \begin{center}
     \includegraphics[width=4.8in,height=2.5in]{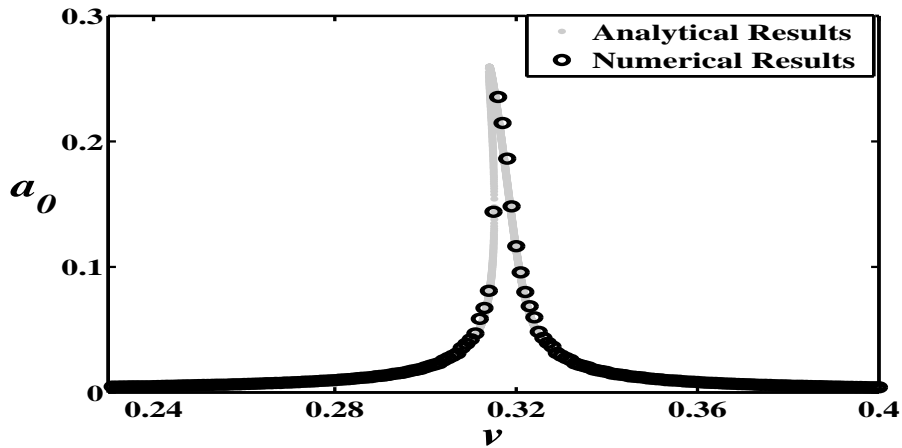}
     \caption{Variations of the steady-state response of deterministic system Eq.~(\ref{eq9}) with $\gamma = 0$ (the other parameters read: $\Gamma = 0.002, N_v = 1, \lambda = 0.0078$): the light gray dots represent the theoretical prediction and the black circles the numerical solution.}
     \label{fig:5}
    \end{center}
    \end{figure*}

\begin{figure}[hbtp]
              \begin{center}
              \includegraphics[width=2.38in,height=2.3in]{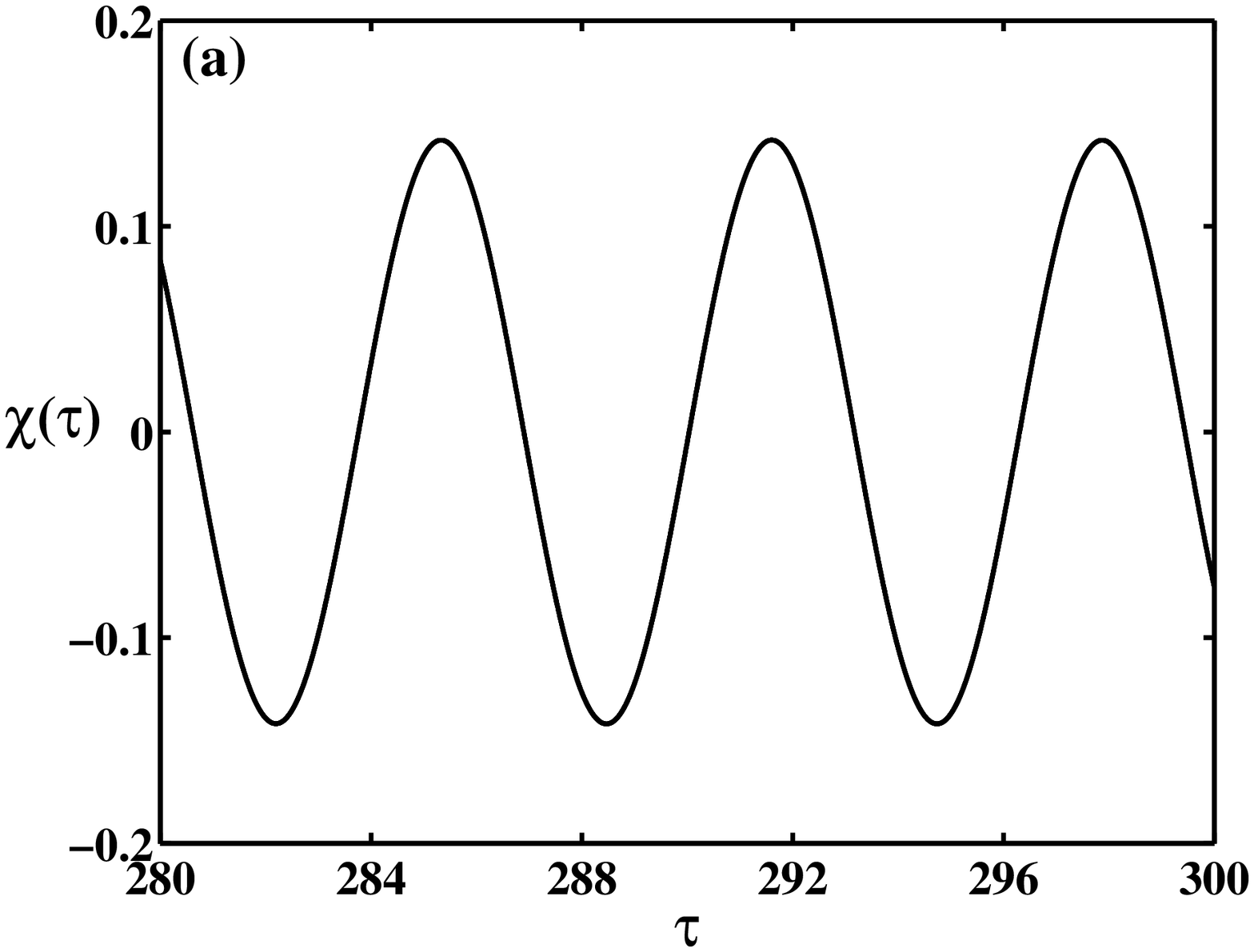}
              \includegraphics[width=2.38in,height=2.3in]{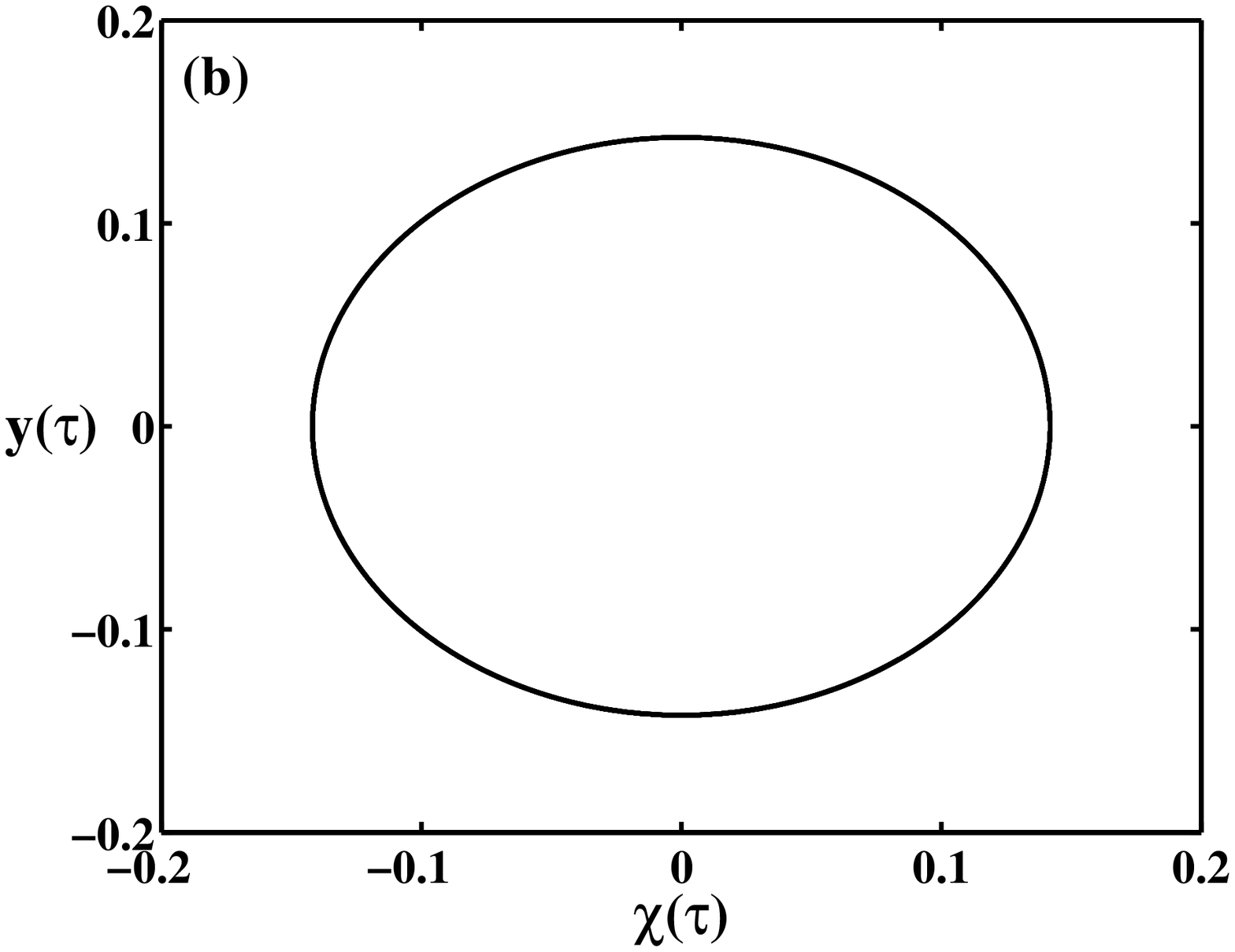}
              \caption{Time evolution and phase portrait of the deterministic system,  Eq.~(\ref{eq9}) with $\gamma = 0$ (the other parameters read: ${\it v} = 0.32, \Gamma = 0.002, N_v = 1, \lambda = 0.0078$): (a) time history of $\chi(\tau)$ and (b) phase plot.}
              \label{fig:6}
             \end{center}
             \end{figure}

In this Section, a Duffing oscillator with bounded Gaussian noise described by  Eq.~(\ref{eq9}) is considered.  In particular, the stationary probability distribution analytically with the method of stochastic averaging method \cite{Stratonovich,Zhu,Roberts,Chamgoue} is sought.
To do so, it is assumed that the noise intensity is small and the following change of variables is introduced,
\begin{equation}\label{eq26}
\begin{array}{l}
\chi (\tau ) = a(\tau )\cos \theta ,\,\,\,\,\,\,\,\,\,\dot \chi (\tau ) =  - \Omega a(\tau )\sin \theta ,  \\
\theta  = \Omega\tau  + \varphi (\tau ).
\end{array}
\end{equation}
Substituting Eqs.~(\ref{eq26}) into Eq.~(\ref{eq9}), we obtain
\begin{equation}\label{eq27}
\left\{ \begin{array}{l}
\dot a =  - \lambda a{\sin ^2}\theta  + \\
\,\,\,\,\,\,\,\,    \frac{1}{\Omega }  \{ a\left( 1 - \Omega ^2 \right)\cos \theta  + \beta a^3 \cos^3\theta  \\
- \Gamma \sin \left[ \Omega \tau  + \gamma W(\tau ) \right] \}\sin \theta \\
\\
a\dot \varphi  =  - \lambda a\sin \theta \cos \theta  + \\
\,\,\,\,\,\,\,\,    \frac{1}{\Omega }  \{a\left( {1 - {\Omega ^2}} \right)\cos \theta  + \beta {a^3}{{\cos }^3}\theta  \\
- \Gamma \sin \left[ {\Omega \tau  + \gamma W(\tau )}  \right]  \}\cos \theta. \,
\end{array} \right.
\end{equation}
The steady-state response for the case of perfect periodicity is considered first, $\gamma= 0$, and Eqs.~(\ref{eq27}) become
\begin{equation}\label{eq28}
\left\{ \begin{array}{l}
\dot a =  - \lambda a{\sin ^2}\theta  + \\
\,\,   \frac{1}{\Omega }\left\{a\left( {1 - {\Omega ^2}} \right)\cos \theta  + \beta {a^3}\cos ^3\theta  - \Gamma \sin \left[ {\theta  - \varphi } \right] \right\}\sin \theta    \\
\, \\
a\dot \varphi  =  - \lambda a\sin \theta \cos \theta  + \\
\,\,  \frac{1}{\Omega }\left\{a\left( {1 - {\Omega ^2}} \right)\cos \theta  + \beta {a^3}{{\cos }^3}\theta  - \Gamma \sin \left( {\theta  - \varphi } \right) \right]\} \cos \theta.    
\end{array} \right  .
\end{equation}
By applying the standard averaging method \cite{Nayfeh}, Eqs.~(\ref{eq28}) reduce to
\begin{equation}\label{eq29}
\left\{ \begin{array}{l}
\dot a =  - \frac{{\lambda a}}{2} - \frac{{\Gamma }}{{2\Omega }}\cos \varphi \\
\\
a\dot \varphi  = \frac{1}{\Omega }\left[ {\frac{{a\left( {1 - {\Omega ^2}} \right)}}{2} + \frac{3}{8}\beta {a^3}} \right] + \frac{{\Gamma }}{{2\Omega }}\sin \varphi .
\end{array} \right.
\end{equation}
The steady-state solutions of Eqs.~(\ref{eq29}) can be found by putting ${\it a = a_0}, \varphi = \varphi_0$ and $\dot a =0$, $\dot \varphi = 0$, this leads to the following result:
\begin{equation}\label{eq30}
\frac{9}{{16}}{\beta ^2}a_0^6 + \frac{3}{2}\beta \left( {1 - {\Omega ^2}} \right)a_0^4 + \left[ {{\lambda ^2}{\Omega ^2} + {{\left( {1 - {\Omega ^2}} \right)}^2}} \right]a_0^2 =  {\Gamma }^2.
\end{equation}
This equation has more than one steady-state solution for some parameters.
The variation of steady-state response ${\it a_0}$ as a function of the speed $v$ is compared with the numerical simulation of Eq.~(\ref{eq9}) and shown in Fig.~\ref{fig:5}.
It can be seen from this figure that the deterministic response predicted by the standard averaging method is in good agreement with that obtained by the numerical simulations.
In particular the resonance is correctly captured by the analysis.
The time response of the system (\ref{eq9}) and phase plot are shown in Fig.~\ref{fig:6} for the noiseless case $\gamma = 0$, and for $ v = 0.32$, $N_v = 1$, $\lambda = 0.0078$.
Clearly, the response is periodic and the phase trajectory is a limit cycle.
 %%%%%%%%%%%%%%%%%%%% FIGURE steady-state amplitude a_0 vs v %%%%%%%%%%%%%%%%%%%%%%%
    %%%%%%%%%%%%%%%%%%%%%%%%%%%%%%%%%%%%%%%%%%%

  %%%%%%%%% FIGURE response time history %%%%%%%%%%%%%%%%%%%
    %%%%%%%%%%%%%%%%%%%%%%%%%%%%%%%%%%%%%%%%%%
Next, the stationary response of system (\ref{eq9}) in the noisy case is determined, $\gamma\neq 0$. To do so, Eqs.~(\ref{eq27}) are rewritten as:
\begin{equation}\label{eq31}
\left\{ \begin{array}{l}
\dot a =  - \lambda a{\sin ^2}\theta \\
 + \frac{1}{\Omega }\left[ {a\left( {1 - {\Omega ^2}} \right)\cos \theta  + \beta {a^3}{{\cos }^3}\theta  - \eta (\tau )} \right]\sin \theta,       \\
\\
\dot \varphi  =  - \lambda \sin \theta \cos \theta  +\\
 \frac{1}{{a\Omega }}\left[ {a\left( {1 - {\Omega ^2}} \right)\cos \theta  + \beta {a^3}{{\cos }^3}\theta  - \eta (\tau )} \right]\cos \theta .
\end{array} \right.
\end{equation}
To apply the method \cite{Stratonovich,Zhu,Roberts,Chamgoue}, one average at the frequency $\Omega$ and the following pair of stochastic equations for amplitude {\it a($\tau$)} and the phase $\varphi(\tau)$ is obtained:
\begin{subequations}
\begin{align}
\label{eq32a}
%\begin{array}{r}
da = \left[ { - \frac{1}{2}\lambda a + \frac{{{{\left( {\Gamma \gamma } \right)}^2}}}{{8a{\Omega ^2}}}\frac{{2{\Omega ^2} + \frac{{{\gamma ^4}}}{4}}}{{{{\left( {\frac{{{\gamma ^4}}}{4}} \right)}^2} + {\Omega ^2}{\gamma ^4}}}} \right]d\tau  +  \nonumber
\\
\frac{{\Gamma \gamma }}{{2\Omega }}\sqrt {\frac{{2{\Omega ^2} + \frac{{{\gamma ^4}}}{4}}}{{{{\left( {\frac{{{\gamma ^4}}}{4}} \right)}^2} + {\Omega ^2}{\gamma ^4}}}} d{W_1}(\tau )
%\end{array}
\\
\label{eq32b}
%\begin{array}{r}
d\varphi  = \frac{1}{\Omega }\left[ {\frac{{\left( {1 - {\Omega ^2}} \right)}}{2} + \frac{3}{8}\beta {a^2}} \right]d\tau  +  \nonumber  \\
 \frac{{\Gamma \gamma }}{{2a\Omega }}\sqrt {\frac{{2{\Omega ^2} + \frac{{{\gamma ^4}}}{4}}}{{{{\left( {\frac{{{\gamma ^4}}}{4}} \right)}^2} + {\Omega ^2}{\gamma ^4}}}} d{W_2}(\tau )
%\end{array}
\end{align}
\label{eq32}
\end{subequations}

Here $W_1(\tau )$ and $W_2(\tau )$ are independent normalized Weiner processes. Clearly, Eq.~(\ref{eq32a}) governing the amplitude $a(\tau)$ does not depend on $\varphi(\tau)$; thus, one can consider the probability density for $a(\tau)$, rather than a joint density for $a(\tau)$ and $\varphi(\tau)$. The probability density function $P(a(\tau),\tau |a(\tau_0),\tau_0)$ for the amplitude is governed by the Fokker-Planck-Kolmogorov equation:
\begin{equation}\label{eq33}
\begin{array}{l}
\frac{{\partial P\left( {a,\tau } \right)}}{{\partial \tau }} = \frac{\partial }{{\partial a}}\left[ {\left( { - \frac{1}{2}\lambda a + \frac{{{{\left( {\Gamma \gamma } \right)}^2}}}{{8a{\Omega ^2}}}\frac{{2{\Omega ^2} + \frac{{{\gamma ^4}}}{4}}}{{{{\left( {\frac{{{\gamma ^4}}}{4}} \right)}^2} + {\Omega ^2}{\gamma ^4}}}} \right)P\left( {a,\tau } \right)} \right]\\
\\
\,\,\,\,\,\,\,\,\,\,\,\,\,\,\,\,\,\,\,\,\,\,\,\,\,\,\,\, + \frac{1}{2}\left( {\frac{{{{\left( {\Gamma \gamma } \right)}^2}}}{{4{\Omega ^2}}}\frac{{2{\Omega ^2} + \frac{{{\gamma ^4}}}{4}}}{{{{\left( {\frac{{{\gamma ^4}}}{4}} \right)}^2} + {\Omega ^2}{\gamma ^4}}}} \right)\frac{{{\partial ^2}P\left( {a,\tau } \right)}}{{\partial {a^2}}}
\end{array}
\end{equation}

In the stationary case, $P:  {{\partial P\left( {a,\tau} \right)}}/{{\partial \tau}} = 0$,  the solution of Eq.~(\ref{eq33}) is:

\begin{equation}\label{eq34}
 P(a) = \frac{Na}{{\varLambda}}\exp \left[ {2\int {\frac{{\Delta (a)}}{{\varLambda}}da} } \right],
  \end{equation}

 \noindent where

 \begin{subequations}
\begin{align}
 \label{eq35a}
%\begin{array}{l}
 \Delta (a) =  - \frac{1}{2}\lambda a + \frac{{{{\left( {\Gamma\gamma } \right)}^2}}}{{8a{\Omega ^2}}}\frac{{2{\Omega ^2} + \frac{{{\gamma ^4}}}{4}}}{{{{\left( {\frac{{{\gamma ^4}}}{4}} \right)}^2} + {\Omega ^2}{\gamma ^4}}},
 \\
\label{eq35b}
\varLambda  = \frac{{{{\left( {\Gamma \gamma } \right)}^2}}}{{4{\Omega ^2}}}\frac{{2{\Omega ^2} + \frac{{{\gamma ^4}}}{4}}}{{{{\left( {\frac{{{\gamma ^4}}}{4}} \right)}^2} + {\Omega ^2}{\gamma ^4}}}.
%\end{array}
\end{align}
\label{eq35}
 \end{subequations}

  Combining Eq.~(\ref{eq34}) and Eqs.~(\ref{eq35}), we get
   \begin{equation}\label{eq36}
P(a) = \frac{\lambda a}{\varLambda}\exp \left(  - \frac{\lambda}{2\varLambda}a^2 \right)
   \end{equation}
%with
%  \begin{equation*}
% \alpha  = \frac{{2\delta {\Omega ^2}\left[ {{{\left( {\frac{{{\gamma ^4}}}{4}} \right)}^2} + {\Omega ^2}{\gamma ^4}} \right]}}{{{{\left( {\Gamma {N_v}\gamma } \right)}^2}\left( {2 {\Omega ^2} + \frac{{{\gamma ^4}}}{4}} \right)}}
%    \end{equation*}

 \noindent
where  $N$ has been determined by the normalization condition:
   \begin{equation}\label{eq37}
   \int_0^\infty  {P(a)da \equiv 1},
   \end{equation}
%that gives the value
%      \begin{equation}\label{eq38}
%  N = 2a \varLambda =\lambda
%   \end{equation}

 %%%%%%%% FIGURE  stationary probability distribution %%%%%%%%%%%%%%%
     %%%%%%%%% FIGURE  stationary probability distribution  %%%%%%%%%%%%%%%%%%%
          \begin{figure}[hbtp]
                    \begin{center}
                      \includegraphics[width=2.38in,height=2.3in]{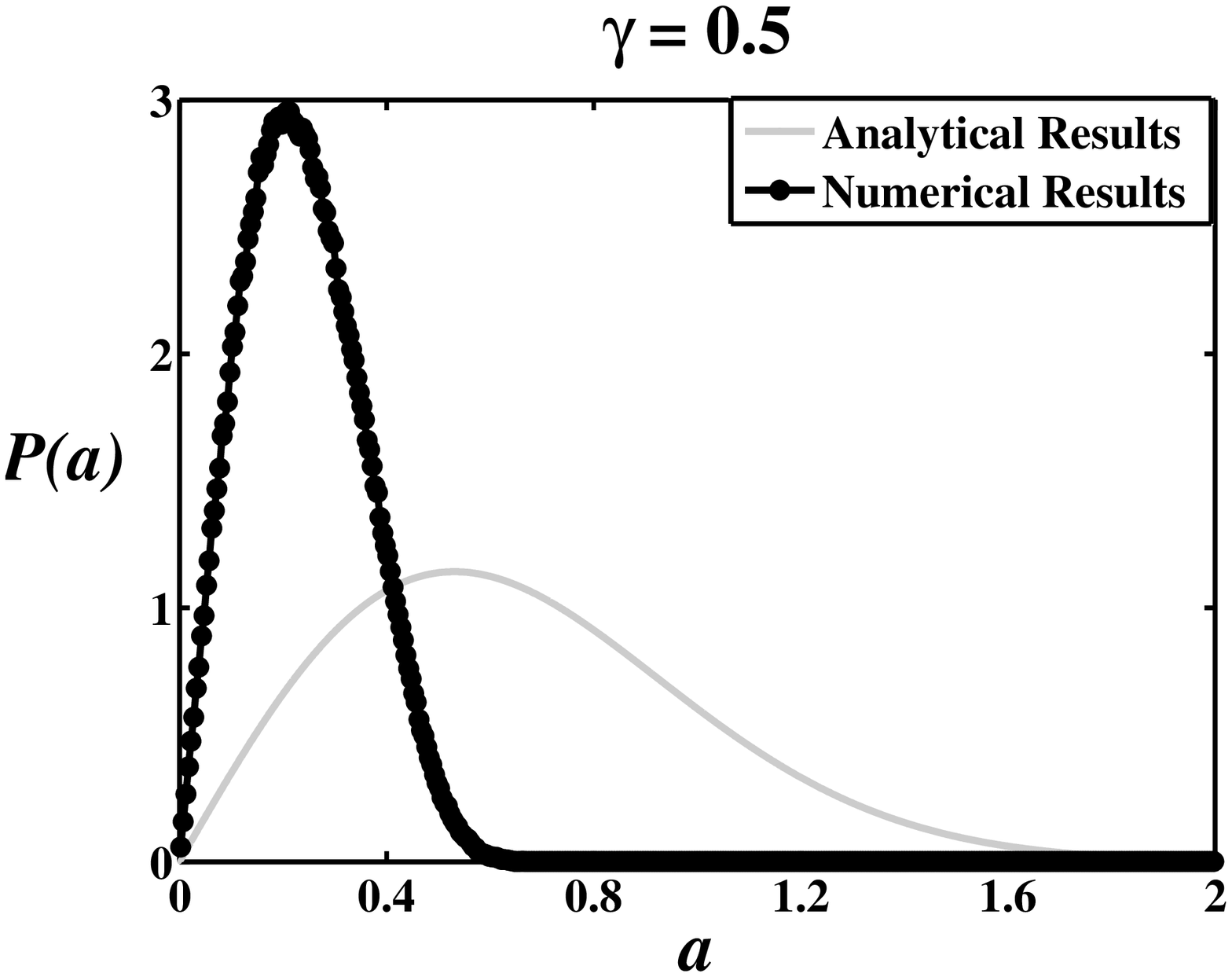}
                      \includegraphics[width=2.38in,height=2.3in]{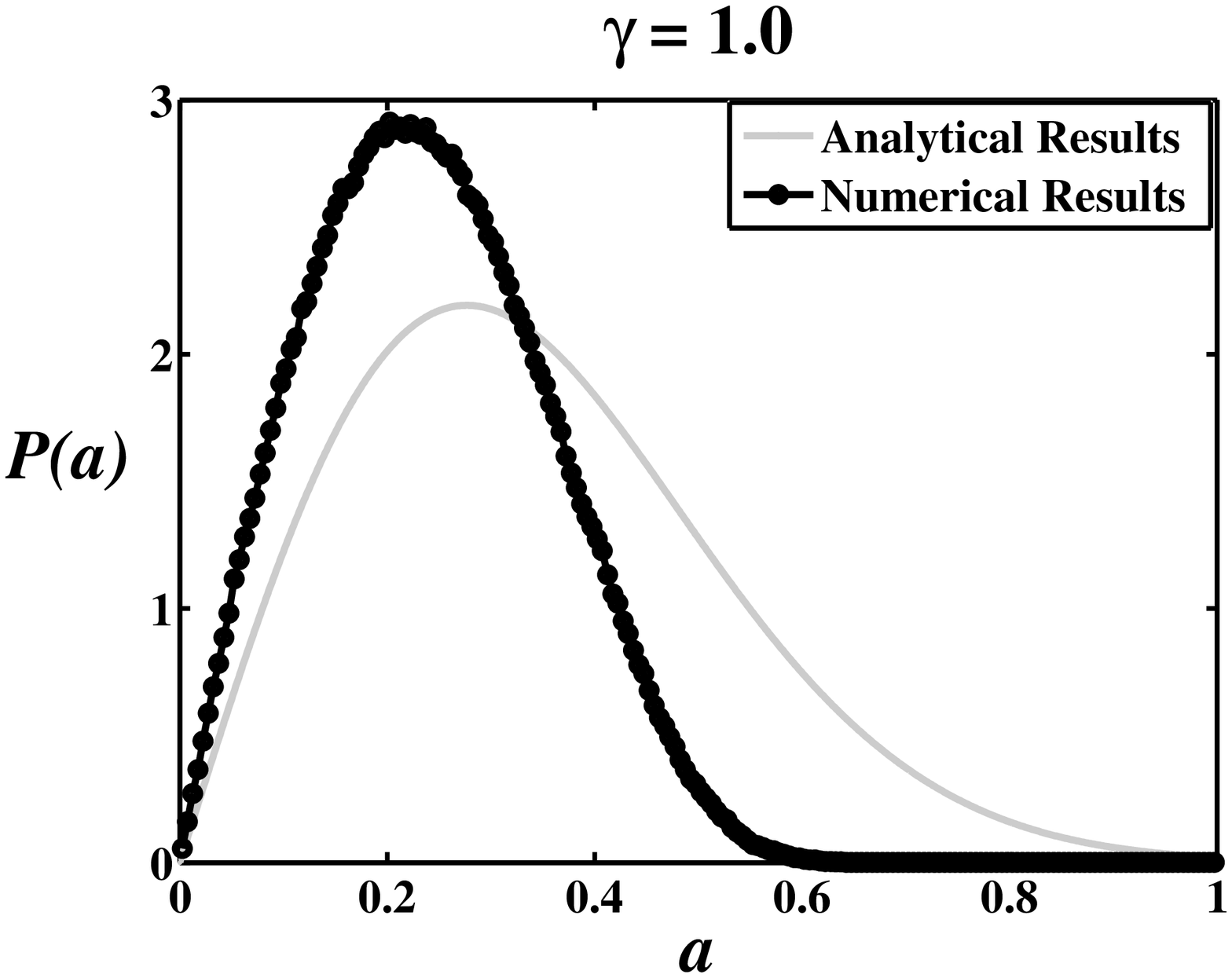}
                   \includegraphics[width=2.38in,height=2.3in]{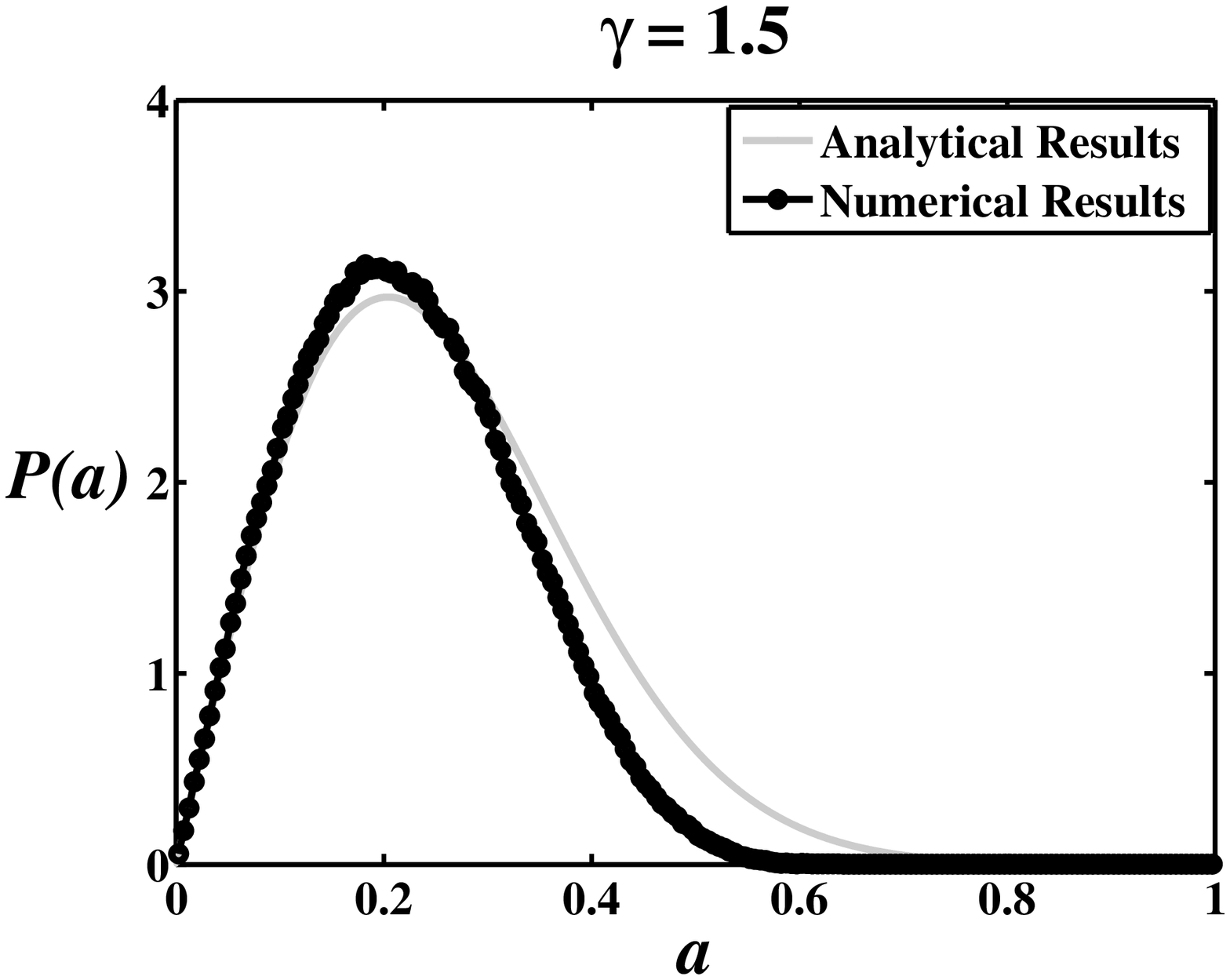}
                       \includegraphics[width=2.38in,height=2.3in]{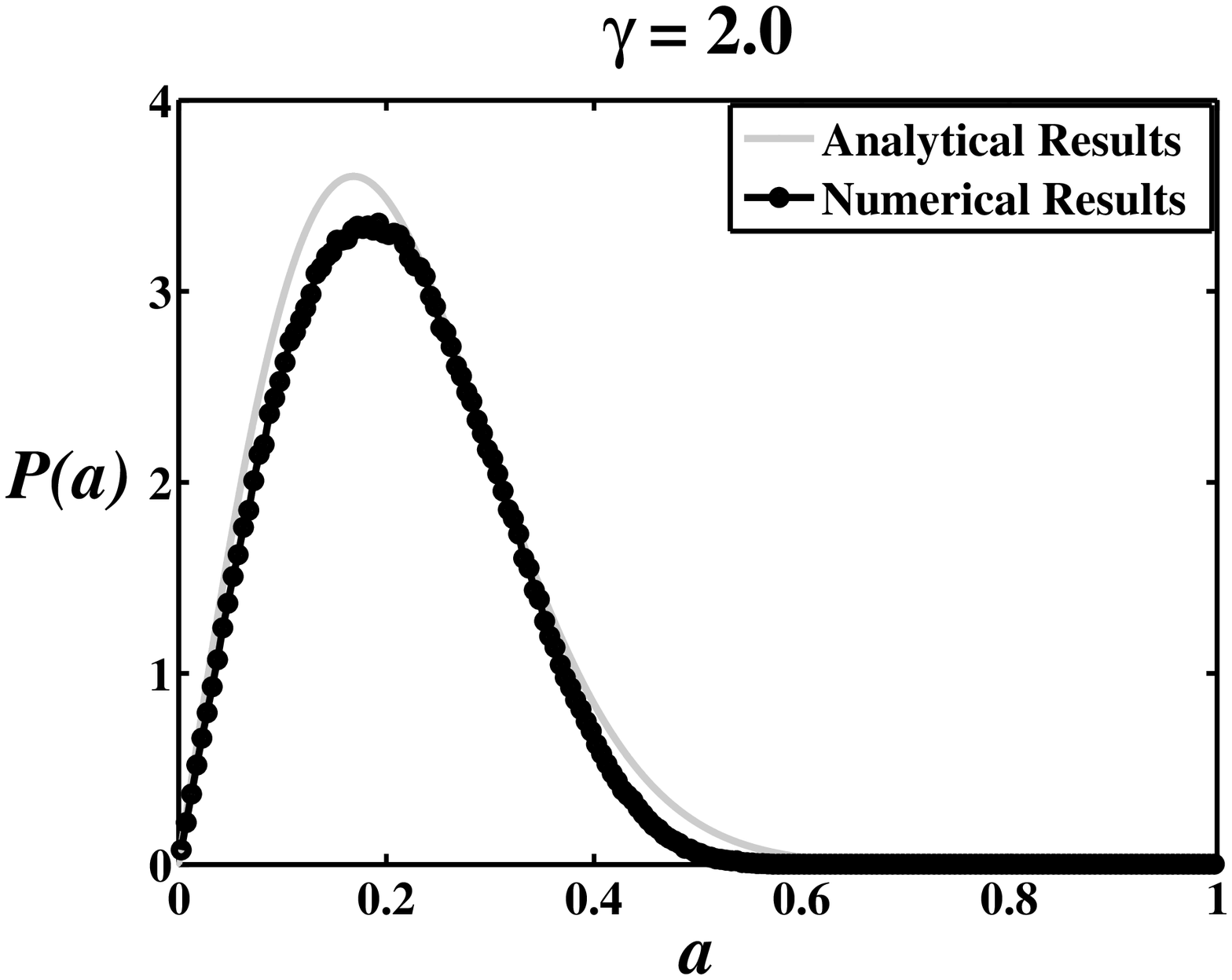}
           \caption{Stationary probability distributions for different values of the standard deviation of stochastic velocity $\gamma$ versus the amplitude ${\it a}$. The parameters used are: $v=0.25$, $\Gamma =0.2$, $\lambda=0.28$, $N_v=1$. The curves with solid light gray lines denote the algebraic calculations using  Eq.~(\ref{eq36}) and the curves with solid black lines (with circles) represent the numerical solutions for the oscillator, Eq.~(\ref{eq9}).}
\label{fig:7}
               \end{center}
            \end{figure}\par
  %%%%%%%%%%%%%%%%%%%%%%%%%%%%%%%%%%%%%%%%%%

\noindent
Moreover, one can find the peaks of the distribution Eq.~(\ref{eq36}) in the points where ${\partial P\left( {a} \right)} / {\partial a} = 0$:

  \begin{equation}\label{eq39}
    a_{m} = \sqrt {\frac{\varLambda}{\lambda} }
  \end{equation}

       \noindent
As the amplitude of the oscillations in the ansatz Eq.~(\ref{eq26}) is defined positive, it is concluded that, the distribution should exhibit just a peak.
In Fig.~\ref{fig:7} the stationary probability density as a function of the amplitude $a$ for different noisy velocity intensity $\gamma$ is displayed, compared with numerical solutions of Eq.~(\ref{eq9}).
The numerical results confirm that the distribution has only one maximum.
 It is also evident that the analytic solution predicts a  peak of the probability distribution at higher values than the numerical solution.
The discrepancy, however, decreases as noise increases, and the agreement is fairly good when fluctuations are very large (compared to the average speed $v=0.25$).
We therefore conclude that there is a better agreement for a two way bridge.

The results obtained in this section confirm that an increase of the noise intensity $\gamma$ leads to a decrease of the amplitude of the oscillations, Eq.(\ref{eq39}).
This is in qualitative agreement with the findings of the Fourier analysis of Section \ref{analysis}, as it confirms the counter-intuitive role of noise, that tends to decrease the amplitude of the oscillations. However, it is underlined that the stochastic averaging does not capture the amplitude of the oscillations at low noise values.

\section{Concluding remarks}
 \label{sec:conclusion}
An analytic approach to the dynamics of a Rayleigh beam under the effect of loads moving with stochastic speed has been considered.
To do so, one has first reduced the full partial differential equation to an effective nonlinear one-dimensional equation.
For this reduced equation, the random terms have been modelled as equal weights moving with a stochastic velocity, thus neglecting other effects as the spread of the weights.
Within these limitations, it has been investigated how the main parameters of the moving loads affect the beam response, and especially how the intensity of the random component of the loads velocity affects the dynamic behaviour of the beam.
Applying the Fourier transform and the theory of residues a mean amplitude equation has been obtained. It has been found that velocity noise first causes an increase of the mean-square amplitude of the beam oscillations, and after a maximum a further increase of the noise causes a {\it decrease} of the mean-square displacement, a behavior that closely follows stochastic resonance phenomena \cite{Gammaitoni98,Wellens04}.
In fact, high levels of noise appear to be paradoxically beneficial for the beam safety, inasmuch the amplitude only increases up to a special noise value where it exhibits a maximum damage to the beam wearing. Such counterintuitive effect of the random disturbance has been extensively debated in signal detection \cite{Ando01} and biological systems \cite{Mcdonnell11}, and are generally connected with suboptimal detection observable as the observed amplitude of the oscillations \cite{Addesso12}.
However, at variance with electronic signal detection where stochastic resonance is extensively studied and can be possibly exploited, the beneficial effect of noise is of course of difficult application in engineering, for the addition of further noise sources is unpractical, if not just impossible.
We notice, however, that the results indicate that negative velocities, corresponding to a two way bridge, might prove safer than a one way beam.
From the physical point of view one ascribe the above behavior to the change in the harmonic content of the bounded noise that is altered, in a nontrivial manner, by an increase of the stochastic intensity.

The analysis has also allowed to estimate the response of the beam to the loads and the other system parameters.
It has been thus demonstrated that the number of moving loads considerably influences the mean-square vibration amplitude of the beam and the appearance of multiple amplitude solutions for some range of the loads weight.
Some numerical simulations of the modal equation, obtained with a stochastic version of the fourth-order Runge-Kutta algorithm, have confirmed the analytic results.

Finally, the stationary probability density function of the amplitude for the noisy modal equation has been analytically derived using a stochastic averaging method and compared to the numerical solution.
It has been found that the agreement is fairly good, and that also this approach confirms that the most probable position to find the oscillator decreases when the intensity of the stochastic fluctuations of the velocity increase. However, stochastic averaging seems to be not able to capture the initial increase of the oscillations amplitude.

\hg{
By way of short summary, we have tried to retrieve what is the effect of the load random velocities on the behavior of the mean amplitude versus the parameters. Some nonlinear dependences are new but not fully unexpected: the number of vehicles changes the resonant velocity (Figs. \ref{fig:2},\ref{fig:5}). More interesting is the unexpected influence of the load, that is not just nonlinear, but reentrant (Fig. \ref{fig:4}). To single out what we regard as the most relevant finding, we focus on the non-monotonic effect of the noise intensity, Fig. \ref{fig:3}, that discloses an unexpected beneficial role of noise intensity on the bridge stability.
}

We would like to conclude mentioning some open issues. First, it could prove interesting to understand the discrepancy between the stochastic averaging approach and the Fourier Transform approach. Second, the noise effect here considered, only in terms of velocity fluctuations, could be complemented with an analysis of other random sources, as the individual weights of the loads. Third, the deterministic underlying equation could possibly be treated with another approach to improve the deterministic approximation.
\begin{acknowledgements}
 G.F. acknowledges partial financial support from MIUR through PON R\&C (Programma Operativo Nazionale "Ricerca e Competitivit\`a") 2007-2013 under Grant No. PON NAFASSY, PONa3\_00007. Part of this work was completed during a research visit of Prof Nana Nbendjo at the University of Kassel in Germany. He is grateful to the Alexander von Humboldt Foundation for financial support within the Georg Forster Fellowship.
\end{acknowledgements}

% BibTeX users please use one of
%\bibliographystyle{spbasic}      % basic style, author-year citations
%\bibliographystyle{spmpsci}      % mathematics and physical sciences
%\bibliographystyle{spphys}       % APS-like style for physics
%\bibliography{}   % name your BibTeX data base

% Non-BibTeX users please use

\end{document}